\documentclass[aps,prl,twocolumn,superscriptaddress,notitlepage,nofootinbib,longbibliography]{revtex4-1}
\usepackage{mathrsfs,color}
\usepackage{booktabs}
\usepackage{float}
\usepackage{tabularx}
\usepackage{epsfig,float}
\usepackage{graphicx}
\usepackage{amsfonts}
\usepackage[figuresright]{rotating}
\usepackage{amssymb}
\usepackage{amsmath}
\usepackage{dcolumn}
\usepackage{bm}
\usepackage{comment}
\usepackage{xcolor}
\usepackage{color}
\usepackage{braket}
\usepackage{units}
\usepackage{xspace}

\usepackage{bbold}
\definecolor{mypink3}{cmyk}{0, 0.7808, 0.4429, 0.1412}
\definecolor{mypink1}{rgb}{0.858, 0.188, 0.478}
\definecolor{mypink2}{RGB}{219, 48, 122}
\usepackage[colorlinks=true, allcolors=mypink1]{hyperref}
\usepackage[fleqn]{mathtools}

\usepackage[shortlabels]{enumitem}
\newcommand{\CSRC}{Beijing Computational Science Research Center, Beijing 100193, China}
\newcommand{\SEU}{Key Laboratory of Quantum Materials and Devices of Ministry of Education, School of Physics, Southeast University, Nanjing 211189, China}
\newcommand{\CNENU}{Center for Quantum Sciences and School of Physics, Northeast Normal University, Changchun 130024, China}
\newcommand{\NENU}{Center for Advanced Optoelectronic Functional Materials Research, and Key Laboratory for UV-Emitting Materials and Technology of Ministry of Education, Northeast Normal University, Changchun 130024, China}

\begin{document}
	
\title{Non-Hermitian-enhanced quantum sensing in an optical interferometer}
	
\author{Xiaojian Huang}
\thanks{These authors contributed equally to this work}
\affiliation{\SEU}
\affiliation{\CSRC}
\author{Lei Xiao}
\thanks{These authors contributed equally to this work}
\email{\\Contact author: xiaoleiphys@seu.edu.cn}
\affiliation{\SEU}
\author{Bingzi Huo}
\affiliation{\CSRC}
\author{X. X. Yi}
\email{Contact author: yixx@nenu.edu.cn}
\affiliation{\CNENU}
\affiliation{\NENU}
\author{Peng Xue}
\affiliation{\SEU}

\begin{abstract}
		The precision of quantum parameter estimation is traditionally constrained by the quantum Cram\'{e}r-Rao bound, which is based on the Hermitian measurement framework. Recent studies of non-Hermitian systems have suggested new possibilities for enhancing parameter-estimation sensitivity. Here, we experimentally realize quantum parameter estimation using a non-Hermitian observable on a linear optical platform. The parameter is encoded in single-photon probe states and read out with a Sagnac interferometer, which allows us to reconstruct the complex expectation value of the implemented non-Hermitian observable from interference fringes. We observe a reduced error-propagation variance compared with the optimal Hermitian observable for the same probe-state model. This advantage remains visible under amplitude-damping noise. We further analyze the complete optical measurement as a physical positive-operator-valued measure (POVM) and show, through the corresponding classical Fisher information (CFI), that the observed non-Hermitian advantage is consistent with the standard quantum metrological limit when all output ports are included. Our results provide an experimental route to non-Hermitian observable readout and clarify its operational meaning in quantum sensing.
\end{abstract}

\maketitle

 \emph{Introduction}---The fundamental limit on the precision of measuring unknown parameters in quantum systems arises from the uncertainty principle of quantum mechanics \cite{Helstrom1969,PhysRevA.87.032324,Giovannetti2004,Helstrom1976,Bao2020b,Holevo1982}. In the framework of quantum parameter estimation theory, measurement precision is typically quantified by the minimum variance of an estimator extracted from a given quantum state. The optimal measurement precision is constrained by the quantum Cram\'{e}r-Rao bound (QCRB) \cite{PhysRevLett.134.010804,helstrom1973cramer,PhysRevResearch.4.013075,PhysRevLett.123.200503}, which is associated with quantum Fisher information (QFI)~\cite{liu2020quantum,PhysRevLett.127.260501,PhysRevResearch.3.043122,PhysRevA.41.4265}. One of the core challenges in quantum metrology is how to design measurement strategies that extract the available parameter information more efficiently \cite{giovannetti2011advances,PhysRevA.96.012117,PhysRevA.102.032607}. Traditionally, quantum observables are represented by Hermitian operators to ensure real-valued outcomes and direct experimental accessibility. Recent developments in non-Hermitian physics have motivated alternative sensing and readout strategies, including non-Hermitian quantum parameter estimation \cite{sciadv_YU,PhysRevA.108.022215,PhysRevLett.124.020501,PhysRevLett.125.240506}, high-precision quantum sensing with exceptional points (EPs) \cite{PhysRevLett.112.203901,Ruan2025,Hodaei2017,Chen2017,Mao2024,PhysRevLett.132.243601,Xue26} and without EPs \cite{PhysRevLett.133.180801,PhysRevLett.124.020501}, and non-Hermitian topological sensing \cite{PhysRevResearch.4.013113,PhysRevLett.125.180403,Parto2025}.
	
	In this work, we experimentally investigate the readout of non-Hermitian observables constructed following Ref.~\cite{li2023enhanced}. Compared with previous works that rely on non-Hermitian evolution to encode the parameter to be estimated \cite{PhysRevLett.133.090801}, our experiment directly implements a non-Hermitian-observable readout using single photons and a Sagnac interferometer. By analyzing the interference patterns, we extract the complex expectation value of the implemented non-Hermitian observable and observe a reduced error-propagation variance compared with the optimal Hermitian observable for the same probe-state model.
\begin{figure}[hbtp]
\centering
\includegraphics[width=\linewidth]{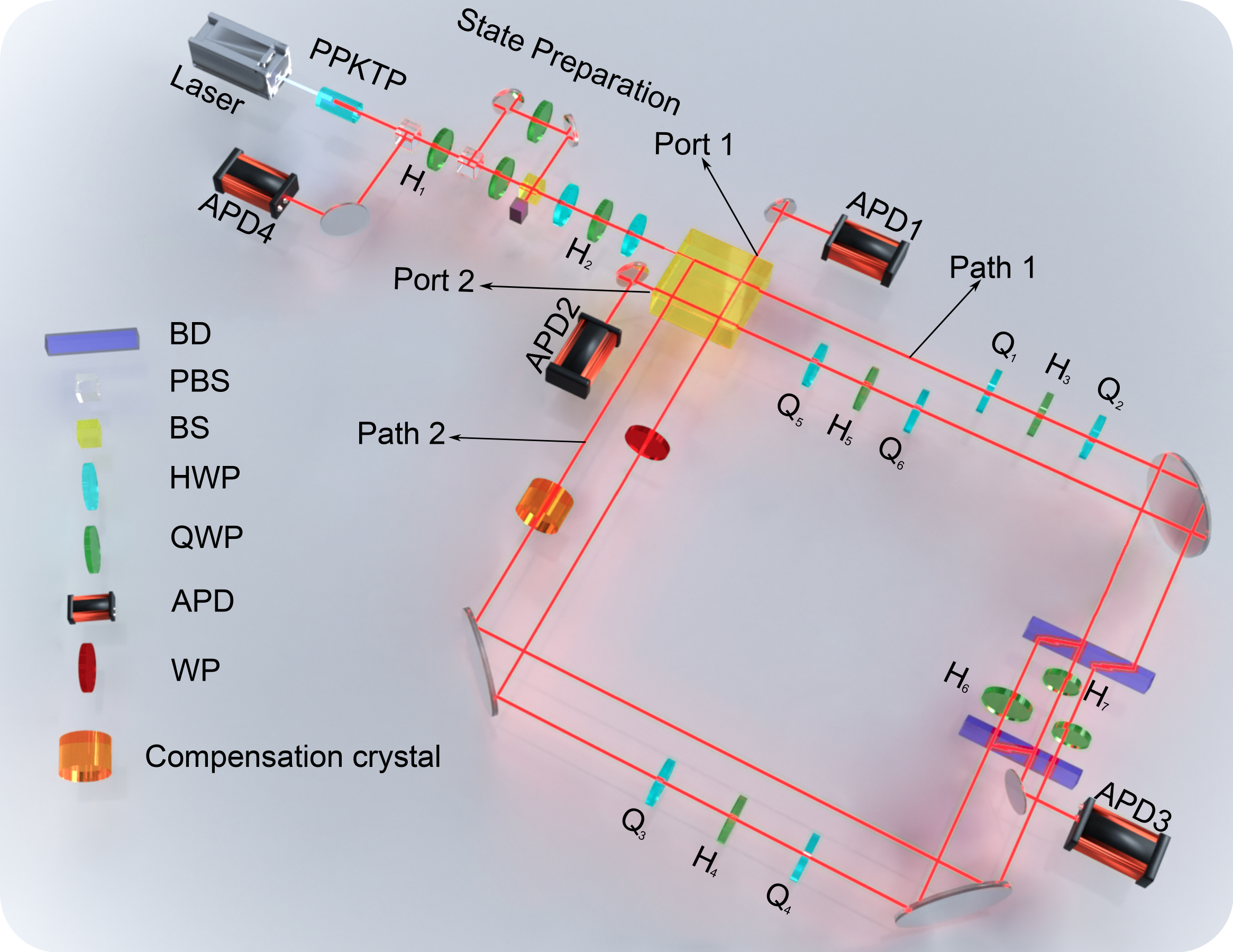}
\caption{Experimental setup. Photon pairs are generated via spontaneous parametric down-conversion. In each pair, one photon serves as the trigger and the other is sent into the state-preparation stage followed by a Sagnac interferometer. In the interferometer, a BS splits the photon into path-$1$ and path-$2$ with equal probability. Along path-$1$, the photon undergoes the unitary evolution $U^\dagger$, while along path-$2$, it undergoes the normalized positive operation $R^\prime$. The two paths propagate in clockwise and counterclockwise directions, respectively, forming a closed loop and interfering at the BS. Coincidence detection of output photons with trigger photons is performed using APDs with a $3$-ns coincidence window.}
\label{setup}
\end{figure}

\emph{ Theoretical framework}---The estimation of a parameter $\theta$ typically involves four main steps: (1) preparing the probe state; (2) encoding the parameter $\theta$ into the probe state; (3) constructing an observable $A$ and measuring its expectation value $\langle A \rangle(\theta)$ for the probe state; and (4) estimating $\theta$ from the measured $\langle A \rangle(\theta)$.

Without loss of generality, we illustrate this process using a single-qubit example. The initial state is described by the density matrix
\begin{equation}
\rho_0=p\ket{\Psi_0}\bra{\Psi_0}+(1-p)\ket{\Psi_1}\bra{\Psi_1},\label{initial}
\end{equation}
where $\ket{\Psi_{0,1}}=\frac{1}{\sqrt{2}}(\ket{0}\pm\ket{1})$, and $p$ denotes the probability that the system is in the state $\ket{\Psi_0}$. We encode the parameter $\theta$ into the initial state $\rho_0$ by a unitary operation $M_\theta=e^{i\theta \sigma_z/2}$, where $\sigma_z$ is the Pauli matrix. The encoding process is given by
\begin{equation}
\rho(\theta )=M_\theta\rho_0M_\theta^{\dagger}.
\label{encode}
\end{equation}
For a non-Hermitian observable $A$, the expectation value is given by $\langle A\rangle(\theta)={\operatorname{Tr}}[\rho(\theta)A]$. The variance of $\theta$ is given by
\begin{equation}
(\Delta \theta)^2=\frac{\left\langle\Delta A^{\dagger} \Delta A\right\rangle}{\partial_\theta\left\langle A^{\dagger}\right\rangle  \partial_\theta\langle A\rangle},\label{va}
\end{equation}
where $\Delta A=A-\langle A \rangle$  \cite{li2023enhanced}.
Therefore, the QCRB of $(\Delta \theta)^2$ in a non-Hermitian system is given by \cite{li2023enhanced}
\begin{equation}
(\Delta\theta)^2\geq\frac{1}{F_\text{nH}},
\label{nonHCR}
\end{equation}
where the non-Hermitian Fisher information is defined as
$F_\text{nH}=\operatorname{Tr}[\rho L^\dagger L]$~\cite{li2023enhanced}.
Here, $L$ is a non-Hermitian logarithmic derivative satisfying
$\partial_\theta\rho=L\rho=\rho L^\dagger$, and is not necessarily Hermitian in general~\cite{li2023enhanced}. The bound is saturated if and only if the observable $A$ is   optimally chosen. Under this condition, one form of the optimal non-Hermitian observable depends on the probe state (in Eq.~\ref{encode}) and  is given by \cite{li2023enhanced}.
\begin{equation}
A_\text{nH}(\theta)=\left(\begin{array}{cc}
\frac{1}{2(-1+p)}+\frac{1}{2 p} & \frac{e^{i \theta}}{2 p^2-2 p} \\
\frac{e^{-i \theta}}{2 p-2 p^2} & \frac{1}{2-2 p}-\frac{1}{2 p}
\end{array}\right).
\label{optnh}
\end{equation}
With this optimal non-Hermitian observable, the QFI is $F_\text{nH}=\frac{(2p-1)^2}{4p(1 - p)}$~\cite{li2023enhanced}. Correspondingly, the QFI for the Hermitian case  is $F_\text{H}=1-4p(1-p)$, with an optimal observable $A_\text{H}(\theta)=\begin{pmatrix} 0 & -ie^{i\theta}\frac{2p-1}{2} \\
ie^{-i\theta}\frac{2p-1}{2} & 0
\end{pmatrix}$, which saturates the traditional QCRB~\cite{li2023enhanced}.

\emph{Experimental Implementation}---For a non-Hermitian observable $A$, the expectation value can be expressed as
$\langle A \rangle=|\langle A\rangle|e^{i\xi}$,
where $|\langle A\rangle|$ is its modulus and
$\xi=\arg(\langle A \rangle)$ is its phase. Direct measurement of non-Hermitian observables remains a critical challenge, as standard quantum measurement excludes non-Hermitian operators. In this work, we construct a Sagnac interferometer \cite{PhysRevLett.76.3053,PhysRevLett.120.230402,PhysRevLett.132.070203} to directly measure the expectation value of a non-Hermitian observable by determining the maximum and minimum intensities of the interference fringes and their corresponding phase shifts. Specifically, we decompose $A$ as $A=UR$~\cite{nirala2019measuring}, where $U=A/\sqrt{A^\dagger A}$ is unitary and $R=\sqrt{A^\dagger A}$ is a Hermitian operator that can be diagonalized as \begin{equation}
\begin{aligned}
	R&=a|\psi_1\rangle\langle\psi_1|+b|\psi_2\rangle\langle\psi_2|
	=a(|\psi_1\rangle\langle\psi_1|+\frac{b}{a}|\psi_2\rangle\langle\psi_2|)\\&=aR^\prime.
\end{aligned}
\end{equation}
Experimentally, we implement the normalized operator $R^\prime$ rather than $R$ itself. Assuming $a>b$, $R^\prime$ can be realized using a beam displacer (BD) and wave plates, with the details provided in the Supplemental Material~\cite{sm}.

As depicted in Fig.~\ref{setup}, we design a Sagnac interferometer in which photons are split into two paths. In path-$1$, the photon undergoes unitary evolution $U^\dagger$, while in path-$2$, it undergoes operator $R^\prime$. Subsequently, the two paths are recombined to interfere, resulting in the formation of an interference pattern. The photon count $I$ received by the detector at port 1 is
\begin{equation}
I=\mathcal{C}[1+\langle R^{\prime2} \rangle +2|\langle A^\prime\rangle|\cos(\xi-\phi+\frac{\pi}{2})],
\label{I}
\end{equation}
where $A=aA^\prime$, $\mathcal{C}=n_0/4$, and $n_0$ is the photon count when $U^\dagger=R^\prime=\hat{I}$. The details of Eq.~(\ref{I}) can be found in the Supplemental Material~\cite{sm}. The maximum and minimum photon counts are $I_{\rm max}=\mathcal{C}(1+\langle R^{\prime2} \rangle +2|\langle A^\prime\rangle|)$ and $I_{\rm min}=\mathcal{C}(1+\langle R^{\prime2} \rangle -2|\langle A^\prime\rangle|)$. Thus, $|\langle A^\prime\rangle|$ and $\langle R^{\prime2}\rangle$ can be directly obtained as
\begin{equation}
	\begin{aligned}
		\langle R^{\prime2}\rangle&=\frac{I_{\rm max}+I_{\rm min}}{2\mathcal{C}}-1,\\
		|\langle A^\prime\rangle|&=\frac{I_{\rm max}-I_{\rm min}}{4\mathcal{C}}.
	\end{aligned}
	\label{RA}
\end{equation}
The expectation value of $A^\prime$ is then
\begin{equation}
	\begin{aligned}
		\langle A^\prime\rangle(\theta)
		&=\frac{I_{\rm max}-I_{\rm min}}{4\mathcal{C}}\cos(\xi)
		+i\frac{I_{\rm max}-I_{\rm min}}{4\mathcal{C}}\sin(\xi)\\
		&={\rm Re}[\langle A^\prime\rangle(\theta)]
		+i{\rm Im}[\langle A^\prime\rangle(\theta)].
	\end{aligned}
	\label{ma}
\end{equation}
The variance of the implemented non-Hermitian observable $A^\prime$ is
\begin{equation}
	\begin{aligned}
		\langle \Delta A^{\prime\dagger}  \Delta A^\prime\rangle
		&=\langle (A^{\prime\dagger}-\langle A^{\prime\dagger}\rangle)
		(A^\prime-\langle A^\prime\rangle)\rangle\\
		&=\langle R^{\prime2} \rangle - |\langle A^\prime \rangle|^2.
	\end{aligned}
	\label{v}
\end{equation}
Since $A=aA^\prime$ differs from $A^\prime$ only by a nonzero scalar factor, the error-propagation variance in Eq.~(\ref{va}) is invariant under this rescaling. Therefore, the experimentally reconstructed $\langle A^\prime\rangle$ and $\langle \Delta A^{\prime\dagger}\Delta A^\prime\rangle$ are sufficient to obtain the same inferred $(\Delta\theta)^2$ as that for $A$.

 To experimentally realize the measurement above, we encode the qubit using the polarization degree of freedom of single photons, with $|0\rangle \leftrightarrow |H\rangle$ and $|1\rangle \leftrightarrow |V\rangle$, where $|H\rangle$ and $|V\rangle$ represent horizontal and vertical polarizations, respectively. The experimental setup is shown in Fig.~\ref{setup}. We generate photon pairs by pumping a 20 mm type-\uppercase\expandafter{\romannumeral2} periodically poled potassium titanyl phosphate (PPKTP) crystal with a $405$ nm continuous-wave diode laser via phase-matched  collinear  spontaneous parametric down-conversion (SPDC) \cite{Shukhin_2015,Kaneda:16,XWD+25,PhysRevA.110.023728}. In each pair, one photon acts as the trigger ($\ket{V}$), while the other serves as the signal ($\ket{H}$). We prepare the initial state using an unbalanced-arm Mach-Zehnder interferometer (MZI)~\cite{PhysRevA.110.053717,Ruan2022,PhysRevLett.127.026404,PhysRevLett.65.1348}, which consists of three half-wave plates (HWPs), two mirrors, a polarizing beam splitter (PBS), and a $50:50$ beam splitter (BS). Polarization control is achieved using HWPs and a quarter-wave plate (QWP). Specifically, the polarization of the signal  photons ($\ket{H}$) is modulated by the HWP ($H_1$) and then split into two paths by the PBS, as shown in Fig.~\ref{setup}. The parameter $p$ of the initial state is determined by the rotation angle $\alpha$ of $H_1$, according to $p=\cos^2 \alpha$. Then, the PBS splits the photons into two paths based on their polarization, with counts $I_1 \propto \sin^2 \alpha$ (path-$1$) and $I_2 \propto \cos^2 \alpha$ (path-$2$). The states $|\Psi_0\rangle$ and $|\Psi_1\rangle$ in Eq.~(\ref{initial}) are prepared by setting the angles of HWPs in two paths to $-22.5^\circ$ and $22.5^\circ$, respectively. Recombining the two paths with two mirrors and a BS yields a state of arbitrary purity, tunable by adjusting $\alpha$ in $[0, \pi/2]$. We then employ a combination of two QWPs at $45^\circ$ and one HWP ($H_2$) at $\frac{\theta+\pi}{4}$, to encode the parameter $\theta$ into the initial state, which corresponds to $M_{\theta}$ in Eq.~(\ref{encode}).

The main part of the device is a phase-adjustable Sagnac interferometer. As depicted in Fig.~\ref{setup}, the BS splits photons into two paths: one transmitted through the BS (path-$1$) and the other reflected by the BS (path-$2$). Photons in both paths are reflected by three mirrors and interfere at the same BS. The unitary operation $U^\dagger$ is applied in path-$1$ using a combination of wave plates $Q_5-H_5-Q_6$, which can achieve any single-qubit unitary operation on polarization of photons~\cite{unitary operation}.
	
Since our experimental setup does not support gain implementation, we realize $R^\prime$ instead of $R$, with the relation $R=aR^\prime$. Details on the construction of $R^\prime$ and its relation to $R$ can be found in the Supplemental Material~\cite{sm}. The Hermitian operator $R^\prime$ is implemented in path-$2$ using an MZI composed of two beam displacers (BDs) and wave plates. We realize $R^\prime$ by introducing photon loss in the $\ket{\psi_2}$ component through the following steps. First, we use the wave-plate group $Q_1-H_3-Q_2$ to transform the polarization basis
($\ket{\psi_1}$ and $\ket{\psi_2}$) in path-$2$ to horizontal ($\ket{H}$) and vertical
($\ket{V}$) states. Then we use two BDs and one HWP ($H_7$) to induce photon loss.
The first BD separates photons in path-$2$ based on their polarization, i.e., the horizontal
polarization component ($\ket{H}$) into the upper path and the vertical polarization component
($\ket{V}$) into the lower path. The $H_6$ transforms the $\ket{H}$ component to $\ket{V}$ and $H_7$ transforms the majority of the $\ket{V}$ component to $\ket{H}$ so that they can be recombined by the other BD. Finally, photons in the unconverted $\ket{V}$ component in the lower path exit the interferometer, thereby introducing a controllable, polarization-dependent photon loss. The angle $\alpha_7$ of $H_7$ is $\alpha_7=\arcsin(b/a)/2$. Finally, we use $Q_3-H_4-Q_4$ to revert the polarization basis back to $\ket{\psi_1}$ and $\ket{\psi_2}$. Details of this construction are provided in the Supplemental Material~\cite{sm}. The phase shift $e^{i\phi}$ is implemented in path-$1$, using a window plate (WP). The photons in the interferometer are finally detected by  avalanche photodiodes (APDs) labeled APD1, APD2 and APD3, in coincidence with the trigger photon detected by APD4. The corresponding counts are denoted as $N_1$, $N_2$ and $N_3$. By tilting the WP, we adjust the relative phase between the two paths to record the interference pattern $I-\phi$.
\begin{figure*}[ht]
\centering
\includegraphics[width=0.8\linewidth]{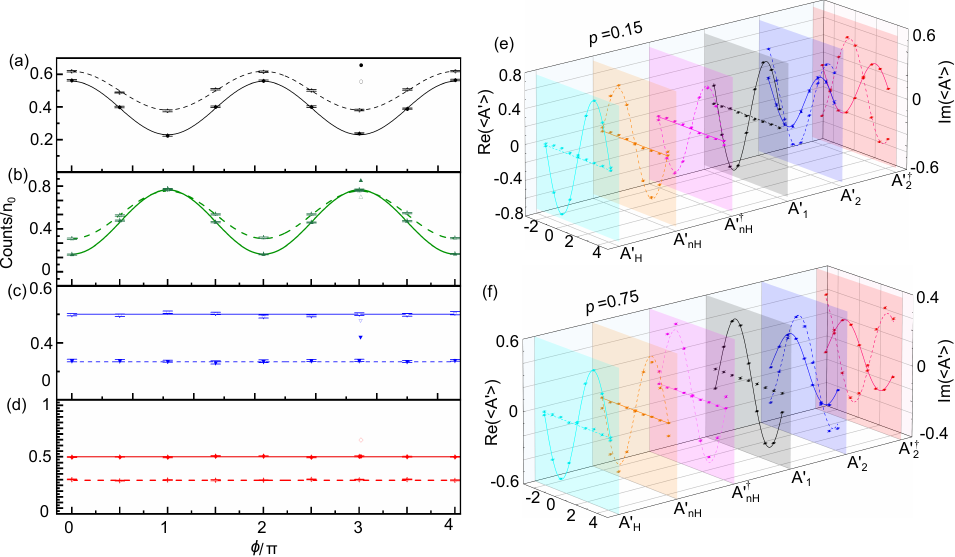}
\caption{ Panels (a)--(d) show photon counts versus $\phi$: (a) The $I(\phi)$ pattern of quantum parameter estimation using the non-optimal non-Hermitian observable $A_1^\prime$ (black solid line) and non-optimal Hermitian observable $A_2^\prime$ (black dashed line) with the initial state $\rho_0$ at $p=0.75$. (b) The $I(\phi)$ pattern of quantum parameter estimation using the non-optimal non-Hermitian observable $A_1^\prime$ (green solid line) and non-optimal Hermitian observable $A_2^\prime$ (green dashed line) with the initial state $\rho_0$ at $p=0.15$. (c) The $I(\phi)$ pattern of the optimal non-Hermitian observable $A^\prime_\text{nH}$ (blue dashed line) and optimal Hermitian observable $A^\prime_\text{H}$ (blue solid line) with the initial state at $p=0.75$. (d) The $I(\phi)$ pattern of the optimal non-Hermitian observable $A^\prime_\text{nH}$ (red dashed line) and optimal Hermitian observable $A^\prime_\text{H}$ (red solid line) with the initial state at $p=0.15$. (e) The real (solid lines) and imaginary (dashed lines) parts of the expectation values for $A_1^\prime$, $A_2^\prime$, $A_2^{\prime\dagger}$, $A^\prime_{\text{H}}$, $A^\prime_{\text{nH}}$, and $A_{\text{nH}}^{\prime\dagger}$, with the initial state at $p=0.15$. (f)  The real (solid lines) and imaginary (dashed lines) parts of the expectation values for $A_1^\prime$, $A_2^\prime$, $A_2^{\prime\dagger}$, $A^\prime_{\text{H}}$, $A^\prime_{\text{nH}}$, and $A_{\text{nH}}^{\prime\dagger}$, with the initial state at $p=0.75$. Error bars are due to the statistical uncertainty in photon number-counting.}
\label{fig:in}
\end{figure*}

{\it Results}---In this section, we present eight representative interference patterns for two different initial states. Figure~\ref{fig:in}(a) presents the $I(\phi)$ pattern of quantum parameter estimation using the non-optimal non-Hermitian observable $A_1$ and non-optimal Hermitian observable $A_2$ with the initial state $\rho_0$ at $p=0.75$, while Fig.~\ref{fig:in}(b) shows the corresponding results at $p=0.15$. Similarly, Fig.~\ref{fig:in}(c) displays the $I(\phi)$ pattern of the optimal non-Hermitian observable $A_\text{nH}$ and optimal Hermitian observable $A_\text{H}$ with the initial state at $p=0.75$, while Fig.~\ref{fig:in}(d) illustrates the $I(\phi)$ pattern at $p=0.15$. The phase $\xi$ can be determined from $\phi(x)$, i.e., $\xi=\frac{\pi}{2}+\phi(x)$. Meanwhile, $\phi$ is obtained by reading out the tilt angle of the window plate (WP). The detailed relationship between $\phi$ and $x$ is provided in the Supplemental Material~\cite{sm}. Hence, the expectation value of observable $A^\prime$ can be obtained through $I_{\rm max}$, $I_{\rm min}$, and $\xi$ according to Eq.~(\ref{ma}). Furthermore, we fix the optimal observable while systematically varying the system states. The real and imaginary parts of these expectation values are then presented in Figs.~\ref{fig:in}(e) and (f) for $p=0.15$ and $p=0.75$, respectively.

\begin{figure*}[htp]
\centering
\includegraphics[width=0.95\linewidth]{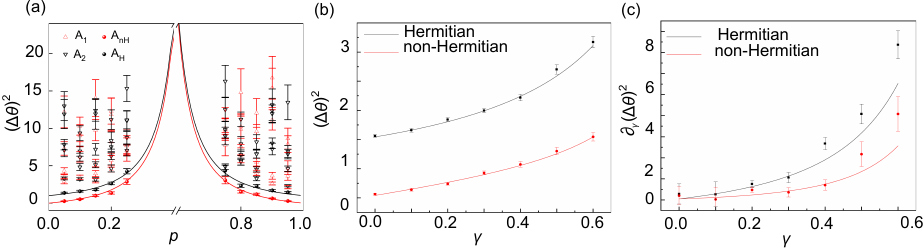}
\caption{(a) Variance $(\Delta\theta)^2$ versus the state parameter $p$ in the noiseless case. Theoretical predictions are shown as lines: red for the optimal non-Hermitian observable and black for the optimal Hermitian observable. Experimental variances are represented by symbols: red circles for the optimal non-Hermitian observable, black circles for the optimal Hermitian observable, red triangles for a non-optimal non-Hermitian observable, and black triangles for a non-optimal Hermitian observable.
	(b) Variance $(\Delta\theta)^2$ versus the amplitude-damping strength $\gamma$ for the optimal Hermitian observable (black) and the optimal non-Hermitian observable (red), with $p=0.01$ and $\theta=0.3\pi$.
	(c) Noise-induced variance change rate $\partial_\gamma(\Delta\theta)^2$ versus $\gamma$ for the same working point.
	Curves represent theoretical predictions, while symbols correspond to experimental results. Error bars are due to the statistical uncertainty in photon number-counting.}
	\label{var}
\end{figure*}

To verify the non-Hermitian QCRB, the derivative of $\langle A^\prime \rangle$ ($\langle A ^{\prime\dagger} \rangle$) with respect to $\theta$ needs to be evaluated according to Eq.~(\ref{nonHCR}). Since we take discretized values of $\theta$, this derivative is numerically estimated according to
\begin{equation}
\begin{aligned}
\frac{\partial \langle A^\prime\rangle}{\partial\theta} &\approx\frac{\langle A^\prime(\theta+\Delta \theta)\rangle-\langle A^\prime(\theta-\Delta\theta)\rangle}{2\Delta \theta},
\end{aligned}
\label{appr}
\end{equation} where the increment is $\Delta\theta=0.1$. We then calculate the variance $(\Delta\theta)^2$ and plot its dependence on $p$ in Fig.~\ref{var}(a). When the optimal observables are selected, both the Hermitian and non-Hermitian cases reach their corresponding error-propagation bounds, while the non-optimal observables give larger variances. The optimal non-Hermitian observable yields a smaller inferred variance than the optimal Hermitian observable over a broad range of $p$, demonstrating non-Hermitian-enhanced parameter estimation at the level of the implemented observable readout.

Because the implementation of $R^\prime$ is intrinsically lossy, the optical device is not merely a measurement of a formal non-Hermitian operator. It is a physical measurement described by a POVM with bright output ports and a loss port. Here, we report the variance inferred from the reconstructed complex expectation value of the non-Hermitian observable. A resource-inclusive analysis based on the classical Fisher information of the complete POVM, including all output ports, is given in the Supplemental Material.

\emph{Noise}---As noise is unavoidable in any measurement process \cite{PhysRevA.97.042112}, we specifically investigate the effect of amplitude damping noise, a common type of decoherence,
	\begin{equation}
		\begin{aligned}
		\varepsilon_{\text{AD}}(\rho) = E_0 \rho E_0^\dagger + E_1 \rho E_1^\dagger,
	\end{aligned}
	\end{equation}
where $E_0 = \begin{pmatrix} 1 & 0 \\
			0 & \sqrt{1 - \gamma}
		\end{pmatrix},
		E_1 = \begin{pmatrix}
			0 & \sqrt{\gamma} \\
			0 & 0
		\end{pmatrix}$. The parameter $\gamma$ $(\in[0,1])$ governs the noise level. The photons pass through the noise channel after the encoding process ($M_\theta$) and before they enter the Sagnac interferometer.  Such a noise channel can be implemented using another Sagnac interferometer~\cite{NJPG}. The details of the noise channel are provided in Supplemental Material~\cite{sm}. To quantify the impact of noise on the error-propagation variance for both Hermitian and non-Hermitian observables, we define the noise-induced variance change rate as,
\begin{equation}
\frac{\partial (\Delta \theta)^2}{\partial \gamma} \approx \frac{(\Delta \theta)^2_{\gamma + \Delta \gamma} - (\Delta \theta)^2_{\gamma - \Delta \gamma}}{2 \Delta \gamma}.
\end{equation}
We choose the initial state with $p=0.01$ and set the parameter to be estimated as $\theta=0.3\pi$. We measure both the variance $(\Delta\theta)^2$ and the noise-induced variance change rate. As shown in Fig.~\ref{var}(b), the variance achieved with the optimal non-Hermitian observable is smaller than that with the optimal Hermitian observable over the measured range of $\gamma$. This indicates a higher readout precision for the implemented non-Hermitian observable under amplitude-damping noise.

The noise-induced variance change rate further clarifies this robustness. For instance, at $\gamma=0.6$, the measured variance change rates for the Hermitian and non-Hermitian observables are $8.3633\pm0.6677$ and $5.0821\pm0.8230$, respectively. This notable difference shows that the non-Hermitian observable is less sensitive to amplitude-damping noise, as shown in Fig.~\ref{var}(c). Moreover, this robustness becomes more pronounced as the noise strength increases. These results demonstrate that, for the selected amplitude-damping channel and working point, the implemented non-Hermitian readout provides a smaller error-propagation variance than the Hermitian benchmark. Our experimental data agree with the theoretical predictions.

\emph{Conclusion and discussion}---{
  We have experimentally demonstrated quantum parameter estimation using an implemented non-Hermitian observable in a single-photon Sagnac interferometer. By reconstructing the complex expectation value of the non-Hermitian observable from interference fringes and applying the error-propagation formula, we observe a smaller inferred variance than that obtained with the optimal Hermitian observable for the same probe-state model. We also show that this readout-level advantage remains visible under amplitude-damping noise.
	
	It is important to distinguish this non-Hermitian readout advantage from a fundamental violation of the standard quantum Fisher-information bound. The optical implementation of $A_\text{nH}$ is a physical POVM rather than an ideal measurement of an abstract non-Hermitian operator. When all output events are retained, including the two bright output ports and the loss port, the relevant figure of merit is the classical Fisher information
	\begin{equation}
		F_{\rm POVM}(\theta)=\sum_j \frac{[\partial_\theta P_j(\theta)]^2}{P_j(\theta)},
	\end{equation}
	where $P_j(\theta)$ are the probabilities of the physical output ports. Our full-POVM analysis shows that this classical Fisher information does not exceed the standard Hermitian optimum. Therefore, the observed enhancement should be understood as a structured non-Hermitian readout advantage, rather than an increase of the total information available from the complete optical measurement. The detailed calculation is provided in the Supplemental Material.
	
	These results clarify the operational meaning of non-Hermitian observable measurements in quantum metrology. They show that non-Hermitian readout can reshape the distribution of information among different measurement channels and provide practical advantages in selected readout configurations, while remaining fully consistent with the standard quantum metrological limits when the complete POVM is considered.
}
	
\begin{acknowledgments}
This work has been supported by the National Key R\&D Program of China (Grant No. 2023YFA1406701), the National Natural Science Foundation of China (Grants No.~92476106, No.~92265209) and the Beijing National Laboratory for Condensed Matter Physics (Grant No.~2024BNLCMPKF010).
\end{acknowledgments}


\begin{thebibliography}{100}
\bibitem{Helstrom1969} C. W. Helstrom, Quantum detection and estimation theory, J. Stat. Phys. {\bf 1}, 231 (1969).
\bibitem{PhysRevA.87.032324} G. T\'{o}th and D. Petz, Extremal properties of the variance and the quantum Fisher information, Phys. Rev. A {\bf 87}, 032324 (2013).
\bibitem{Giovannetti2004} V. Giovannetti, S. Lloyd, and L. Maccone, Quantum-enhanced measurements: beating the standard quantum limit, Science {\bf 306}, 1330--1336 (2004).
\bibitem{Helstrom1976} C. W. Helstrom, \textit{Quantum Detection and Estimation Theory} (Academic Press, New York, 1976).
\bibitem{Bao2020b} H. Bao, S. Jin, J. Duan, S. Jia, K. M{\o}lmer, H. Shen, and Y. Xiao, Retrodiction beyond the Heisenberg uncertainty relation, Nat. Commun. {\bf 11}, 5658 (2020).
\bibitem{Holevo1982} A. S. Holevo, \textit{Probabilistic and Statistical Aspects of Quantum Theory} (North-Holland, Amsterdam, 1982).

\bibitem{PhysRevLett.134.010804} J. R. Hervas, A. Z. Goldberg, A. S. Sanz, Z. Hradil, J. \v{R}eh\'a\v{c}ek, and L. L. S\'anchez-Soto, Beyond the quantum Cram\'er-Rao bound, Phys. Rev. Lett. {\bf 134}, 010804 (2025).
\bibitem{helstrom1973cramer}C. W. Helstrom, Cram\'er-Rao inequalities for operator-valued measures in quantum mechanics, Int. J. Theor. Phys. {\bf 8}, 361 (1973).
\bibitem{PhysRevResearch.4.013075} G. T\'{o}th and F. Fr\"{o}wis, Uncertainty relations with the variance and the quantum Fisher information based on convex decompositions of density matrices, Phys. Rev. Res. {\bf 4}, 013075 (2022).
\bibitem{PhysRevLett.123.200503} F. Albarelli, J. F. Friel, and A. Datta, Evaluating the Holevo Cram\'er-Rao bound for multiparameter quantum metrology, Phys. Rev. Lett. {\bf 123}, 200503 (2019).
\bibitem{liu2020quantum} J. Liu, H. Yuan, X.-M. Lu, and X. Wang, Quantum Fisher information matrix and multiparameter estimation, J. Phys. A: Math. Theor. {\bf 53}, 023001 (2020).

\bibitem{PhysRevResearch.3.043122} M. Yu, D. Li, J. Wang, Y. Chu, P. Yang, M. Gong, N. Goldman, and J. Cai, Experimental estimation of the quantum Fisher information from randomized measurements, Phys. Rev. Res. {\bf 3}, 043122 (2021).
\bibitem{PhysRevA.41.4265} B. R. Frieden, Fisher information, disorder, and the equilibrium distributions of physics,
Phys. Rev. A {\bf 41}, 4265--4276 (1990).
\bibitem{PhysRevLett.127.260501}A. Rath, C. Branciard, A. Minguzzi, and B. Vermersch, Quantum Fisher information from randomized measurements, Phys. Rev. Lett. {\bf 127}, 260501 (2021).

\bibitem{giovannetti2011advances}V. Giovannetti, S. Lloyd, and L. Maccone, Advances in quantum metrology, Nat. Photonics {\bf 5}, 222 (2011).
\bibitem{PhysRevA.96.012117} J. Liu and H. Yuan, Quantum parameter estimation with optimal control, Phys. Rev. A {\bf 96}, 012117 (2017).
\bibitem{PhysRevA.102.032607} W. Wu and C. Shi, Quantum parameter estimation in a dissipative environment, Phys. Rev. A {\bf 102}, 032607 (2020).

\bibitem{PhysRevLett.117.110802} Z.-P. Liu, J. Zhang, \c{S}. K. \"{O}zdemir, B. Peng, H. Jing, X.-Y. L\"{u}, C.-W. Li, L. Yang, F. Nori, and Y.-X. Liu, Metrology with $\mathcal{PT}$-symmetric cavities: enhanced sensitivity near the $\mathcal{PT}$-phase transition, Phys. Rev. Lett. {\bf 117}, 110802 (2016).
\bibitem{Bao2020} H. Bao, J. Duan, S. Jin, X. Lu, P. Li, W. Qu, M. Wang, I. Novikova, E. E. Mikhailov, K.-F. Zhao, K. M{\o}lmer, H. Shen, and Y. Xiao, Spin squeezing of $10^{11}$ atoms by prediction and retrodiction measurements, Nature {\bf 581}, 159--163 (2020).
\bibitem{sciadv_YU} X. Yu, X. Zhao, L. Li, X.-M. Hu, X. Duan, H. Yuan, and C. Zhang, Toward Heisenberg scaling in non-Hermitian metrology at the quantum regime, Sci. Adv. {\bf 10}, eadk7616 (2024).

\bibitem{PhysRevA.108.022215} X. Yu and C. Zhang, Quantum parameter estimation of non-Hermitian systems with optimal measurements, Phys. Rev. A {\bf 108}, 022215 (2023).

\bibitem{PhysRevLett.124.020501} Y. Chu, Y. Liu, H. Liu, and J. Cai, Quantum sensing with a single-qubit pseudo-Hermitian system, Phys. Rev. Lett. {\bf 124}, 020501 (2020).

\bibitem{PhysRevLett.125.240506} S. Yu, Y. Meng, J.-S. Tang, X.-Y. Xu, Y.-T. Wang, P. Yin, Z.-J. Ke, W. Liu, Z.-P. Li, Y.-Z. Yang, G. Chen, Y.-J. Han, C.-F. Li, and G.-C. Guo, Experimental investigation of quantum $\mathcal{PT}$-enhanced sensor, Phys. Rev. Lett. {\bf 125}, 240506 (2020).

\bibitem{PhysRevLett.112.203901} J. Wiersig, Enhancing the sensitivity of frequency and energy splitting detection by using exceptional points: application to microcavity sensors for single-particle detection, Phys. Rev. Lett. {\bf 112}, 203901 (2014).

\bibitem{Ruan2025} Y.-P. Ruan, J.-S. Tang, Z. Li, H. Wu, W. Zhou, L. Xiao, J. Chen, S.-J. Ge, W. Hu, H. Zhang, C.-W. Qiu, W. Liu, H. Jing, Y.-Q. Lu, and K. Xia, Observation of loss-enhanced magneto-optical effect, Nat. Photonics {\bf 19}, 109--115 (2025).

\bibitem{Hodaei2017} H. Hodaei, A. U. Hassan, S. Wittek, H. Garcia-Gracia, R. El-Ganainy, D. N. Christodoulides, and M. Khajavikhan, Enhanced sensitivity at higher-order exceptional points, Nature {\bf 548}, 187--191 (2017).

\bibitem{Chen2017} W. Chen, \c{S}. K. \"{O}zdemir, G. Zhao, J. Wiersig, and L. Yang, Exceptional points enhance sensing in an optical microcavity, Nature {\bf 548}, 192--196 (2017).

\bibitem{Mao2024} W. Mao, Z. Fu, Y. Li, F. Li, and L. Yang, Exceptional-point-enhanced phase sensing, Sci. Adv. {\bf 10}, eadl5037 (2024).

\bibitem{PhysRevLett.132.243601} H. Loughlin and V. Sudhir, Exceptional-point sensors offer no fundamental signal-to-noise ratio enhancement, Phys. Rev. Lett. {\bf 132}, 243601 (2024).

\bibitem{Xue26} P. Xue, Essay: Topological Phases and Exceptional Points in Non-Hermitian Systems, {\it Phys. Rev. Lett.} {\bf 136}, 170001 (2026).

\bibitem{PhysRevLett.133.180801} L. Xiao, Y. Chu, Q. Lin, H. Lin, W. Yi, J. Cai, and P. Xue, Non-Hermitian sensing in the absence of exceptional points, Phys. Rev. Lett. {\bf 133}, 180801 (2024).

\bibitem{PhysRevResearch.4.013113} F. Koch and J. C. Budich, Quantum non-Hermitian topological sensors, Phys. Rev. Res. {\bf 4}, 013113 (2022).

\bibitem{PhysRevLett.125.180403} J. C. Budich and E. J. Bergholtz, Non-Hermitian topological sensors, Phys. Rev. Lett. {\bf 125}, 180403 (2020).

\bibitem{Parto2025} M. Parto, C. Leefmans, J. Williams, R. M. Gray, and A. Marandi, Enhanced sensitivity via non-Hermitian topology, Light: Sci. \& Appl. {\bf 14}, 6 (2025).

\bibitem{li2023enhanced} J. Li, H. Liu, Z. Wang, and X. X. Yi, Enhanced parameter estimation by measurement of non-Hermitian operators, AAPPS Bull. {\bf 33}, 22 (2023).
\bibitem{PhysRevLett.133.090801} J.-X. Peng, B. Zhu, W. Zhang, and K. Zhang, Enhanced quantum metrology with non-phase-covariant noise, Phys. Rev. Lett. {\bf 133}, 090801 (2024).

\bibitem{PhysRevLett.76.3053} K.-X. Sun, M. M. Fejer, E. Gustafson, and R. L. Byer, Sagnac interferometer for gravitational-wave detection, Phys. Rev. Lett. {\bf 76}, 3053 (1996).

\bibitem{PhysRevLett.120.230402}K.-W. Bong, N. Tischler, R. B. Patel, S. Wollmann, G. J. Pryde, and M. J. W. Hall, Strong unitary and overlap uncertainty relations: Theory and experiment, Phys. Rev. Lett. {\bf 120}, 230402 (2018).


\bibitem{PhysRevLett.132.070203} X. Zhao, X. Yu, W. Zhou, C. Zhang, J.-S. Xu, C.-F. Li, and G.-C. Guo, Experimental investigation of uncertainty relations for non-Hermitian operators, Phys. Rev. Lett. {\bf 132}, 070203 (2024).

\bibitem{nirala2019measuring}G. Nirala, S. N. Sahoo, A. K. Pati, and U. Sinha, Measuring average of non-Hermitian operator with weak value in a Mach-Zehnder interferometer, Phys. Rev. A {\bf 99}, 022111 (2019).

\bibitem{sm}See Supplemental Material for details.
\bibitem{Shukhin_2015}A. A. Shukhin, D. O. Akatiev, I. Z. Latypov, A. V. Shkalikov, and A. A. Kalachev, Simulating single-photon sources based on backward-wave spontaneous parametric down-conversion in a periodically poled KTP waveguide, J. Phys. Conf. Ser. {\bf 613}, 012015 (2015).
\bibitem{PhysRevA.110.023728} E. R. Hellebek, K. M{\o}lmer, and A. S. S{\o}rensen, Characterization of the multimode nature of single-photon sources based on spontaneous parametric down-conversion, Phys. Rev. A {\bf 110}, 023728 (2024).
\bibitem{Kaneda:16} F. Kaneda, K. Garay-Palmett, A. B. U'Ren, and P. G. Kwiat, Heralded single-photon source utilizing highly nondegenerate, spectrally factorable spontaneous parametric downconversion, Opt. Express {\bf 24}, 10733--10747 (2016).

\bibitem{XWD+25} L. Xiao, K. Wang, D. Qu, H. Gao, Q. Lin, Z. Bian, X. Zhan, and P. Xue, Non-Hermitian physics in photonic systems, {\it Photonics Insights} {\bf 4(3)}, R09 (2025).

\bibitem{PhysRevLett.127.026404}K. Wang, L. Xiao, J. C. Budich, W. Yi, and P. Xue, Simulating exceptional non-Hermitian metals with single-photon interferometry, Phys. Rev. Lett. {\bf 127}, 026404 (2021).
\bibitem{Ruan2022} Y.-P. Ruan, H.-D. Wu, S.-J. Ge, L. Tang, Z.-X. Li, H. Zhang, F. Xu, W. Hu, M. Xiao, Y.-Q. Lu, and K.-Y. Xia, Ultralow-power all-optical switching via a chiral Mach-Zehnder interferometer, Opt. Express {\bf 30}, 19199--19211 (2022).
\bibitem{PhysRevLett.65.1348} J. G. Rarity, P. R. Tapster, E. Jakeman, T. Larchuk, R. A. Campos, M. C. Teich, and B. E. A. Saleh, Two-photon interference in a Mach-Zehnder interferometer, Phys. Rev. Lett. {\bf 65}, 1348--1351 (1990).
\bibitem{PhysRevA.110.053717}X. Huang, L. Xiao, K. Wang, and P. Xue, Experimental realization of quantum state engineering with single photons, Phys. Rev. A {\bf 110}, 053717 (2024).

\bibitem{unitary operation} {B. N. Simon, C. M. Chandrashekar, and S. Simon, Hamilton's turns as a visual tool kit for designing single-qubit unitary gates, Phys. Rev. A {\bf 85}, 022323 (2012)}.

\bibitem{PhysRevA.97.042112}K. Wang, X. Wang, X. Zhan, Z. Bian, J. Li, B. C. Sanders, and P. Xue, Entanglement-enhanced quantum metrology in a noisy environment, Phys. Rev. A {\bf 97}, 042112 (2018).

\bibitem{NJPG}H. Gao, L. Xiao, K. Wang, D. Qu, Q. Lin, and P. Xue, Experimental verification of trade-off relation for coherence and disturbance, New J. Phys. {\bf 24}, 073011 (2022).

\bibitem{PhysRevLett.131.160801} W. Ding, X. Wang, and S. Chen, Fundamental sensitivity limits for non-Hermitian quantum sensors, Phys. Rev. Lett. {\bf 131}, 160801 (2023).

\bibitem{Zeng2025postselected}
N. Zeng, T. Liu, K. Xia, Y.-R. Zhang, and F. Nori,
Non-Hermitian sensing from the perspective of post-selected measurements, Phys. Rev. Research {\bf 7},  043219  (2025).

\end{thebibliography}
\end{document}